\documentclass[a4paper,12pt]{article}
\usepackage{amsmath}
\begin{document}

\title{Helicity fluctuation, generation of linking number and effect on resistivity}
\author{F. Spineanu, M. Vlad \\
Association EURATOM-MEC Romania \\
NILPRP MG-36, Magurele, Bucharest, Romania \\
spineanu@ifin.nipne.ro}
\maketitle

\begin{abstract}
The energy of the stochastic magnetic field is bounded from below by a
topological quantity expressing the degree of linkage of the field lines.
When the bound is saturated one can assume that the storage of a certain
magnetic energy requires a minimal degree of topological complexity. It is
then possible to infer a connection between the helicity content and the
average curvature of the magnetic field lines. The random curvature induce
random drifts leading to an additional dissipation and modified resistivity.
\end{abstract}

\section{Introduction}

When the Chirikov criterion is verified for several chains of magnetic
islands (developing at closely neighbor resonant magnetic surfaces in a
volume of plasma) the magnetic field becomes stochastic. In general the
magnetic stochasticity is taken into account in transport processes due to
the high efficiency of energy spreading through the stochastic region.
However from the point of view of the structure of the magnetic field it is
difficult to say anything more than we know from Hamiltonian chaotic
systems: there is a stochastic (exponential) instability, local Lyapunov
exponents and Kolmogorov length and the test particles move diffusively or
have various sub- and supra-diffusive behaviors.

However in a stochastic region the field must still obey some constraints.
These constraints arise from the relation between the energy stored in the
magnetic field and the topological complexity of the field. The constraints
can be briefly expressed in this way: it is not possible to support a
certain energy in a volume spanned by transiently stochastic magnetic field
lines if these magnetic field lines do not have a certain minimal degree of
topological complexity.

This should be seen in relation with the equation that expresses the
topological \emph{link }in terms of \emph{writhe} and \emph{twist} and in
relation with the dynamics of a twisted flux tube. If an initial amount of
link is stored exclusively as twist, then beyond a certain level of twist
the flux tube deforms and acquires writhe, thus distributing the higher
amount of link into the two kinds of topological deformations: twist and
writhe. In a plasma free from strong magnetic background (as in
astrophysical plasma or solar corona) generation of writhe means a coiling
or super-coiling instability, a large spatial deformation. In a tokamak the
stochastic flux tubes are also subject to the writhing instability when a
local fluctuation of the parallel electric field occurs, but they are more
constraint by the confining $\mathbf{B}_{0}$ and cannot perform large
spatial displacements. Instead, as a result of small deformations
originating from local writhing (coiling) instability, they will reconnect
such that, from elements of tubes, effectively new strings are created, with
a new effective entanglement. It is reasonable to assume that these new,
episodic, flux tubes, by their mutual linking, satisfy on the average the
energy-topology constraints. Therefore we will assume that the field flux
tubes inside the stochastic region will reconnect to generate transiently
configurations that exhibit a certain topological entanglement. Together
with the dynamical nature of the stochasticity phenomena, the formation of
these entangled structures is transient and we may suppose that the higher
topological content results from a statistical average. At large time the
topological reduction occurs with suppression of relative linking via tube
merging, a process called by Parker \emph{topological dissipation} \cite
{parker}. 

\section{Energy and topology of divergenceless vector fields}

For two curves $\gamma _{1}$ and $\gamma _{2}$, the \emph{link }invariant is
given by the formula (Gauss) 
\begin{equation}
Lk\left( \gamma _{1},\gamma _{2}\right) =\frac{1}{4\pi }\oint_{\gamma
_{1}}dx^{\mu }\oint_{\gamma _{2}}dy^{\nu }\varepsilon _{\mu \nu \rho }\frac{%
\left( x-y\right) ^{\rho }}{\left| x-y\right| ^{3}}  \label{eq1}
\end{equation}
This is an integer number and represents the relative entanglement of two
magnetic lines in the stochastic region. If a line closes in itself (as a
magnetic line on a resonant surface in tokamak), the formula can still be
applied, giving the self-linking. It is obtained by taking $\gamma
_{1}\equiv \gamma $, $\gamma _{2}\equiv \gamma +\varepsilon \widehat{\mathbf{%
n}}$ with $\widehat{\mathbf{n}}$ a versor perpendicular to the tangent of $%
\gamma $, and taking the limit $\varepsilon \rightarrow 0$ (this operation
is called \emph{framing}). However for a flux tube a more complex situation
arises: the magnetic field in the tube has the lines twisted relative to the
axis and the topological description is given in terms of the \emph{twist}
invariant. It is calculated by considering a line on the surface of the tube
and the axis of the tube and defining the vectors: $\mathbf{T}\left(
s\right) $, the tangent to the axis of the tube; $\mathbf{U}\left( s\right) $%
, the versor from the current point on the axis toward the current point on
the line; $s$ is the length along the axis. Then the \emph{twist} is defined
as 
\begin{equation}
Tw=\oint_{\gamma }ds\left[ \mathbf{T}\left( s\right) \times \mathbf{U}\left(
s\right) \right] \cdot \frac{d\mathbf{U}\left( s\right) }{ds}  \label{eq2}
\end{equation}

The deformation of the flux tube of axis $\gamma $ relative to the plane is
measured by the topological number \emph{writhe}, defined as 
\begin{equation}
Wr\left( \gamma \right) =\frac{1}{4\pi }\oint_{\gamma }dx^{\mu
}\oint_{\gamma }dy^{\nu }\varepsilon _{\mu \nu \rho }\frac{\left( x-y\right)
^{\rho }}{\left| x-y\right| ^{3}}  \label{eq3}
\end{equation}
While the \emph{twist} measures the rate at which a line twists around the
axis of the tube, the writhe measures the rate at which the axis of the tube
is twisted in space. The following relation exists between the three basic
topological numbers for a flux tube 
\begin{equation}
Lk=Tw+Wr  \label{eq4}
\end{equation}

Instead of generalizing $Lk$ to an arbitrary but finite number of curves
(magnetic lines) in space, it is defined an equivalent topological quantity
(also noted $Lk$) which refers this time to a vector field in the volume.
Instead of the Gauss link (a discrete set of curves) the definition will now
imply a continuous, field-like invariant, the Chern-Simons action, which is
the total helicity. For a divergenceless vector field $\xi $ (velocity or
magnetic field) in $R^{3}$ the helicity is defined as 
\begin{equation}
H\left( \xi \right) =\int_{M}d^{3}x\left( \xi ,curl^{-1}\xi \right)
\label{eq5}
\end{equation}
which is the same as the integral of $\mathbf{v\cdot \omega }$ or $\mathbf{%
A\cdot B}$ over volume. Consider two narrow, linked flux tubes of the vector
field $\xi $, $\gamma _{1}$ and $\gamma _{2}$. Then 
\begin{equation}
H\left( \xi \right) =2Lk\left( \gamma _{1},\gamma _{2}\right) \cdot \left|
flux_{1}\right| \cdot \left| flux_{2}\right|  \label{eq6}
\end{equation}
which shows that the \emph{link} of the magnetic flux tubes is a measure of
the magnetic \emph{helicity} in the volume. The total helicity in the volume
is the integral of the Chern-Simons form and with adequate boundary
conditions this is an integer number, a direct consequence of the
topological nature of the \emph{link} invariant (invariance refers here to
deformations of the field $\xi \left( \equiv \mathbf{B}\right) $ that do not
break and reconnect the lines) 
\begin{equation}
Q=\frac{1}{32\pi ^{2}}\int d^{3}x\varepsilon ^{jkl}F_{jk}A_{l}  \label{eq7}
\end{equation}
where $A_{l}$ is the magnetic potential and $F_{jk}$ is the electromagnetic
stress tensor. The integrand is the Chern-Simons form, or helicity density,
for the magnetic field. The integer $Q$ is called the Hopf invariant. The
magnetic lines of a field $\mathbf{B}$ or the streamlines of a flow $\mathbf{%
v}$ are obtained form equations like $\sum dx_{i}/B_{i}=0$, generally
difficult to solve. Therefore using these solutions to construct topological
invariants is very difficult and we would need a different representation of
the lines for easier handling. This is provided by the Skyrme-Faddeev model,
or the modified $O\left( 3\right) $ nonlinear sigma model, where a line is a
topological soliton, clearly exhibiting topological properties (see Ward 
\cite{ward}). It is very suggestive that this model has recently been
derived precisely starting from the plasma of electrons and ions, coupled to
electromagnetic field (Faddeev and Niemi \cite{fadniemi}). It is then
legitimate to use the general results derived for the Skyrme-Faddeev model
and in particular the following lower bound for the energy of the magnetic
field. The inequality is 
\begin{equation}
E\geq \eta Q^{3/4}  \label{eq8}
\end{equation}
where $\eta $ is a constant. It means that the energy is bounded from below
by the $3/4$-th power of the total helicity content in the volume or by a
quantity that contains the \emph{total linking} of magnetic lines in the
volume, at the power $3/4$.

A more practical measure of the topological content is the \emph{average
crossing number} $C$, obtained for a pair of lines by summing the signed
intersections in the plane-projection of the spatial curves, averaged over
all directions of projection. It differs of Eq.(\ref{eq3}) by taking the
absolute value of the mixed product. Friedmann and He \cite{friedhe} have
extended the concept for a continuous field. We follow the argument of
Berger \cite{berger} to find the energy bound $E\geq const\;C^{2}$.

Consider magnetic flux tubes whose ends are tied to points situated in two
parallel planes (at distance $L$) and are linked one with the others. Taking
two points on their axis, we connect them with a line and measure the angle
formed by this line with a fixed direction in one of the plane of
projection. This quantity is $\theta _{12}$. The crossing number can be
expressed using this angle 
\begin{equation}
\overline{c}=\frac{1}{\pi }\int_{0}^{L}dz\left| \frac{d\theta _{12}}{dz}%
\right|  \label{eq9}
\end{equation}

The magnetic fields has magnitudes $B_{z1}$ and $B_{z2}$. We combine
energetic and topological quantities by weighting the angle variation along $%
z$ with the two magnetic fluxes 
\begin{equation}
C=\frac{1}{2\pi }\int_{0}^{L}dz\int d^{2}x_{1}\int
d^{2}x_{2}B_{z1}B_{z2}\left| \frac{d\theta _{12}}{dz}\right|  \label{eq10}
\end{equation}

The angle $\theta _{12}$ is generated by the deviation of the lines (with
tangent versors $\widehat{\mathbf{n}}_{1,2}$) relative to a reference
straight vertical line normal to the end planes. This deviation is produced
by the component of the magnetic field which is perpendicular on the main
field $B_{z}$, $d\mathbf{x}_{1}/dz=\mathbf{B}_{\perp 1}\left( \mathbf{x}%
_{1}\right) /B_{0}$. Then one finds 
\begin{equation}
\frac{d\theta _{12}}{dz}=\frac{1}{r_{12}}\left( \widehat{\mathbf{n}}_{2}-%
\widehat{\mathbf{n}}_{1}\right) \cdot \widehat{\mathbf{e}}_{\theta _{12}}
\label{eq11}
\end{equation}
The versor of the line connecting points on the two magnetic field lines is $%
\widehat{\mathbf{e}}_{\theta _{12}}=\widehat{\mathbf{e}}_{z}\times \widehat{%
\mathbf{r}}_{12}$. Then 
\begin{equation}
\frac{dC}{dz}=\int \int d^{2}x_{1}d^{2}x_{2}\frac{B_{z1}B_{z2}}{2\pi r_{12}}%
\left| \left( \widehat{\mathbf{n}}_{2}-\widehat{\mathbf{n}}_{1}\right) \cdot 
\widehat{\mathbf{e}}_{\theta _{12}}\right|  \label{eq12}
\end{equation}

The energy of the magnetic field in a volume is $E_{f}=\frac{B_{z}^{2}}{2\mu
_{0}}\int d^{3}x\mathbf{b}^{2}$ where $\mathbf{b\equiv }\left(
b_{x},b_{y}\right) =\mathbf{B}_{\perp }/B_{z}$. By successive bounds Berger
finds the inequality 
\begin{equation}
E_{f}\geq \text{const}\;C^{2}  \label{eq13}
\end{equation}

We now have two inequalities implying the energy of the magnetic field in a
region and two measures of the topological content in that volume: one is
the \emph{linking} of the magnetic lines, or equivalently, the helicity $H$,
and the other is a more geometrical characterization of the entanglement,
the \emph{average crossing number}, $C$.

Taking the inequalities as saturated we can estimate the total average
crossing number for a certain amount of helicity $H$ in the volume 
\begin{equation}
C\sim H^{3/8}  \label{eq14}
\end{equation}

\bigskip

The next step is to connect the crossing number with geometrical properties
of a generic magnetic field line. For a magnetic field line $\gamma $ the
topological quantity \emph{crossing number} $C$ can be estimated from the
number $\Omega \left( \gamma \right) $ of intersections of the line $\gamma $
with an arbitrary surface. There is a general theorem that allows to
estimate this number as 
\begin{equation}
2\pi \Omega \left( \gamma \right) \sim 4K_{1}+3K_{2}  \label{eq15}
\end{equation}
where $K_{1,2}$ are integrals of Frenet curvatures. Taking $\left| K\right| $
as the upper estimation of the local value of the curvature along $\gamma $,
we have 
\begin{equation}
C\sim \Omega \left( \gamma \right) \sim L\left| K\right|  \label{eq16}
\end{equation}
where $L$ is the length of the tube. Then we have 
\begin{equation}
\left| K\right| \sim H^{3/8}  \label{eq17}
\end{equation}
These very qualitative estimations led us to a scaling law connecting the
average of the local curvature of a typical magnetic field line with the
helicity inside the finite volume of the stochastic region. A magnetic line
is curved since it is linked with other lines, and this link is generated
for the magnetic structure to be able to store the energy in a stochastic
region. Generation of linking also occurs when a certain amount of helicity
is injected in a plasma volume. It is reasonable to assume that the
curvature is distributed randomly in the volume.

\section{Effects of topology on resistivity and diffusion}

The curvature of magnetic flux tubes induces drifts of particles. Electrons
and ions flowing along curved magnetic lines will have opposite drifts and
local charge separations produce random transversal electric fields. For a
finite collisionality this is a source of additional dissipation. The
equation for ions is 
\begin{equation}
ev_{\parallel i}E_{z}\frac{\partial \widetilde{f}_{i}}{\partial \varepsilon }%
+ev_{di}\widetilde{E}_{\perp }\frac{\partial f_{i0}}{\partial \varepsilon }%
=-\nu \widetilde{f}_{i}  \label{eq18}
\end{equation}
for a transversal field $\widetilde{E}_{\perp }\sim \eta _{0}n_{0}\frac{T_{e}%
}{B}\left( 1+\tau \right) \left| K\right| $ and $\left| K\right| =\left|
\left( \widehat{\mathbf{n}}\cdot \mathbf{\nabla }\right) \widehat{\mathbf{n}}%
\right| $, for which we can use an estimation based on Eq.(\ref{eq17}). We
finally obtain an estimation of the negative current perturbation due to
curvature 
\begin{equation}
\left| \delta \widetilde{j}_{\parallel }\right| \sim e\frac{\eta _{0}}{\nu }%
\frac{n_{0}^{2}T_{e}}{B^{2}}\sqrt{\frac{T_{i}}{m_{i}}}\left( 1+\tau \right)
\left| K\right| ^{2}\frac{1+\rho +\rho ^{2}}{1+\rho }  \label{eq19}
\end{equation}
with $\rho \equiv \left( \frac{e_{i}v_{\parallel i}E_{z}}{\nu T_{i}}\right)
^{1/2}$. This is not a substantial modification of the equilibrium current
(less than $1\%$), which means that the enhanced resistivity is mainly due
to other processes.

In general it is assumed that the magnetic reconnection does not affect the
total helicity. However there is a dynamic redistribution of helicity (with
overall conservation) since in the stochastic region there are filaments of
current and local increase of the parallel electric field, from which we
have $dh/dt=-2\mathbf{E\cdot B}$. Then this mechanism is a potential
feedback loop: higher resistivity leads to higher reconnection rate and
higher helicity perturbation, which in turn creates magnetic linking and
curvature.

\bigskip

In an alternative approach to the problem of topology of the magnetic field
in a stochastic region, we can base our estimations on the much simpler
assumption, that a magnetic line is randomly ``scattered'' at equal space ($%
z $) intervals and performs a random walk. Actually this is the classical
assumption for the diffusion in stochastic magnetic fields. To characterize
quantitatively the topology of the line we use the analogy with the polymer
entanglement. A functional integral formalism can be developed (Tanaka \cite
{tanaka}) taking as action the free random walk, with the constraint that
the line has a fixed, $m$, winding number around a reference straight line.
The mean square dispersion of the winding number (linking) can then be
calculated 
\begin{equation}
\left\langle m^{2}\right\rangle \sim \frac{1}{4\pi }\left( \ln N\right) ^{2}
\label{eq20}
\end{equation}
where $N$ is the number of steps. Since $k_{\parallel }$ is low for magnetic
turbulence, the winding number is a small number. But this represents the
random winding naturally occuring in an unconstrained random walk of the
magnetic line, when the magnetic perturbation is a Gaussian noise. Actually,
we know that a given amount of helicity can only be realised by a certain
volume-averaged mutual linking and this is an effective constraint which can
only be realised through a much higher density of $\sqrt{\left\langle
m^{2}\right\rangle }$ than the free random motion. Then the higher winding
leads to a sort of trapping for the magnetic lines and the effective
diffusion will be smaller that for the brownian case \cite{flmadi1}.

\bigskip

In general the use of topological quantities can improve the description of
stochastic magnetic fields, \emph{e.g.} diffusion, Kolmogorov length, etc.
These are usually expressed in terms of mean square amplitude of the
perturbation, $\left\langle \left| \widetilde{b}\right| ^{2}\right\rangle $
but including the topological quantities can lead to more refined models.

Consider the line $\gamma $ of a perturbed magnetic field and the equation $%
Du=0$ where $D$ is the covariant derivative, similar to the velocity, $v=p-A$
which is applied on a function $u$ along the line $\gamma $. The equation
says that the covariant derivative along the line, of the function $u$ is
zero. Then $\left( \frac{\partial }{\partial s}-iA\right) u=0$ leads to $%
u=u_{0}\exp \left( i\oint_{\gamma }dsA\left( s\right) \right) $. It is
natural to make a generalization of the two-dimensional concept of
point-like vortices and introduce the spinors along the magnetic line $%
\gamma $. By the same arguments (Spineanu and Vlad \cite{flmadi}) we will
need the dual (dotted-indices) spinors and we need to represent $A$ in $%
SU\left( 2\right) $. Then, more generally, the expression of $u$ is 
\begin{equation}
W_{\gamma }\equiv \mathrm{Tr}_{R}P\exp \left( i\oint_{\gamma }dx^{\mu
}A_{\mu }\right)  \label{eq21}
\end{equation}
the trace is over the representation $R$ of $SU\left( 2\right) $ and $P$ is
the ordering along $\gamma $. Being a closed path this number is a
functional of $\gamma $ and of $A$. This is the Wilson loop.

We subject the fluctuations of the potential $A$ to the constraint of
minimum helicity in the volume, because lower helicity allows lower energy
according to the bound Eq.(\ref{eq8}). The Boltzmann weight in the partition
function is then the exponential of an \emph{action} representing the total
helicity, \emph{i.e.} the integral of the Chern-Simons density (compared
with Eq.(\ref{eq7}), here $A$ is a matrix) 
\begin{equation}
S=\frac{\kappa }{4\pi }\int_{M^{3}}d^{3}r\varepsilon ^{\mu \nu \rho }\mathrm{%
Tr}\left( A_{\mu }\partial _{\nu }A_{\rho }-\frac{2}{3}A_{\mu }A_{\nu
}A_{\rho }\right)  \label{eq22}
\end{equation}
The average of $W_{\gamma }$ is 
\begin{equation}
\int D\left[ A\right] W_{\gamma }\exp \left( S\right)  \label{eq23}
\end{equation}
can be expanded in powers $\kappa ^{-n}$ . The first significant term (order 
$\kappa ^{-1}$) is the integral of the two-point correlation of the
fluctuating potential 
\begin{equation}
-\mathrm{Tr}\left[ \left( R^{a}R^{b}\right) \oint_{\gamma }dx^{\mu
}\int^{x}dy^{\nu }\left\langle A_{\nu }^{a}\left( y\right) A_{\mu
}^{b}\left( x\right) \right\rangle \right] \sim Wr\left( \gamma \right)
\label{eq24}
\end{equation}
where $Wr\left( \gamma \right) $ is the \emph{writhe} number of the curve $%
\gamma $ ($a,b$ are labels in $SU\left( 2\right) $). Calculating $Lk$%
=self-linking of the curve $\gamma $, the classical relation is obtained
between the \emph{link}, the \emph{twist} and the \emph{writhe }$Lk\left(
\gamma \right) =Tw\left( \gamma \right) +Wr\left( \gamma \right) $.

Therefore if the fluctuation of the poloidal flux function $\psi $, or the $%
z $-component of the magnetic potential $A_{z}$ are such that higher
helicity states are difficult to access, then the two-point correlations of
the perturbed potential along a curve $\gamma $ can be expressed as the
kernel of the Gauss integral for the self-linking number of $\gamma $. The
result 
\begin{equation}
\left\langle \psi \left( x_{1}\right) \psi \left( x_{2}\right) \right\rangle
_{\gamma }\sim \text{integrand of }Lk\left( \gamma \right)  \label{eq25}
\end{equation}
possibly sheds a new light on the correlations of fluctuating magnetic
quantities, since we now express them also by topological quantities.

\section{Discussion}

We have examined the topological constraints on the stochastic magnetic
configuration when a transient increase of helicity occurs in a finite
plasma volume. Via bounds related to the magnetic energy that can be safely
stored in that volume (\emph{i.e.} a statistical stationarity can be
attained) a scaling can be derived between the helicity and the average
curvature of a generic magnetic line in the volume. The particles' 
curvature-drift-induced new dissipation appears to not modify substantially
the resistivity. However the new instruments that imply the topology of
magnetic field are useful : the average dispersion of the winding of a line
relative to a reference axis serves to quantify the trapping of a line and
the reduction of the classical magnetic diffusion.

We note finally that using these analytical instruments the \emph{%
topological dissipation} process may be described by the coupling of the
magnetic helicity density (the Chern-Simons Lagrangian density) with a
pseudoscalar field. The dynamics of this field is that of the kinematic
helicity of the plasma and again a field-theoretical description appears to
be possible.

\end{document}